# SLTR: Simultaneous Localization of Target and Reflector in NLOS Condition Using Beacons

Muhammad.H Fares†′′′′, Hadi Moradi† ‡*, Mahmoud Shahabadi⁂

† Advanced Robotics and Intelligent Systems Laboratory, School of ECE, University of Tehran, Iran. E-mail: m.h.fares@ut.ac.ir

′′′′ EDST, Lebanese University; Hadath, Lebanon.

‡ Advanced Robotics and Intelligent Systems Laboratory, School of ECE, University of Tehran, Iran; Adjunct Research Prof., ISRI, SKKU, South Korea.

* Corresponding author. E-mail: moradih@ut.ac.ir

⁂ School of ECE, University of Tehran, Islamic Republic of Iran. E-mail: shahabad@ut.ac.ir

**Abstract**

When the direct view between the target and the observer is not available, due to obstacles with non-zero sizes, the observation is received after reflection from a reflector, this is the indirect view or Non-Line-Of Sight condition. Localization of a target in NLOS condition still one of the open problems yet. In this paper, we address this problem by localizing the reflector and the target simultaneously using a single stationary receiver, and a determined number of beacons, in which their placements are also analyzed in an unknown map. The work is done in mirror space, when the receiver is a camera, and the reflector is a planar mirror. Furthermore, the distance from the observer to the target is estimated by size constancy concept, and the angle of coming signal is the same as the orientation of the camera, with respect to a global frame. The results show the validation of the proposed work and the simulation results are matched with the theoretical results.

**Key words:** Beacon, Target Localization, Mirror space, NLOS condition, Size constancy concept.

1. Introduction

When most of the authors went in the direction of assuming known map in the workspace to localize a target, and then they proposed beacon placement approaches for these special maps[1, 2], we introduce a new beacon-based approach that is available for localizing the reflector and the target simultaneously in an unknown map, using only one receiver and a determined number of stationary beacons.

Recently, researchers have strived to discover new methods for target localization to perform an accurate target's positioning. Hence, the range-based schemes that have been proposed to deal with the localization problem need the node to node distances to be known, or the angle of the coming signal from different nodes. These needed information are obtained by some methods such as TOA [3], TDOA [4], DOA [5] and RSSI [6] for LOS condition. For NLOS condition, one of the solutions is to combine the mentioned methods, i.e. the DOA, TOA, TDOA and RSSI; in [7], the authors proposed a hybrid TOA and AOA cooperative localization for the NLOS environments in a WSN. Another study is introduced in [8] when Dai et. Al proposed a combination between RSS and AOA measurements to improve the localization accuracy in WSN. Other combinations are shown in [9-11].

Beacons are nodes with accurate positions (for example, equipped with GPS). These nodes play an important role in localizing the reflectors if they are placed in the desired areas, that allow them to be seen by the observer, in order to determine the target's position that depends on the reflector's position and orientation. The beacon-based approach is one of the other methods that have been widely used for the target localization in NLOS [1, 2, 10, 12-18]. However, the beacon's type differs from scenario to another. For example, the authors in [15, 16, 19, 20] proposed the use of a single beacon that moves in a determined path, to ensure the localization of the target. Other studies adopt the use of non-moving beacons, in which their number grows with respect to the searching areas[1, 12, 13, 15].

The work is addressed for different spaces, assuming the different signal models, and the different noises and measurements. The concept of the work remains the same. Such signals are the sound signals [21, 22] that can be received by an array of microphones, the RF signals from mobile phones used in case of search and rescue missions [23, 24] and the light coming from an object and seen by a camera [25, 26]. From the last example (i.e. the object seen by the camera), the distance from the camera to the object can be calculated based on a reference distance for a reference size, in other words, using the size constancy concept, the distance from the observer to the target can be estimated based on its size shown at the image plane of the camera. Furthermore, in a world frame, the orientation of the coming observation is identical to the orientation of the camera. For the sake of simplicity, the experimental work is done in a mirror space, in which the observer is a camera, the reflector is a planar mirror and the target must be localized is an object with known size constancy value. The noises are modeled for this space, and added for the overall work. The detailed work is shown in Fig.1; a lot of possible positions of the target can be seen, where the angle and orientation of the reflector are unknown, for one-bounce reflection. Here, the beacon's position affects the localization and helps for finding the reflector's components and the target's position.

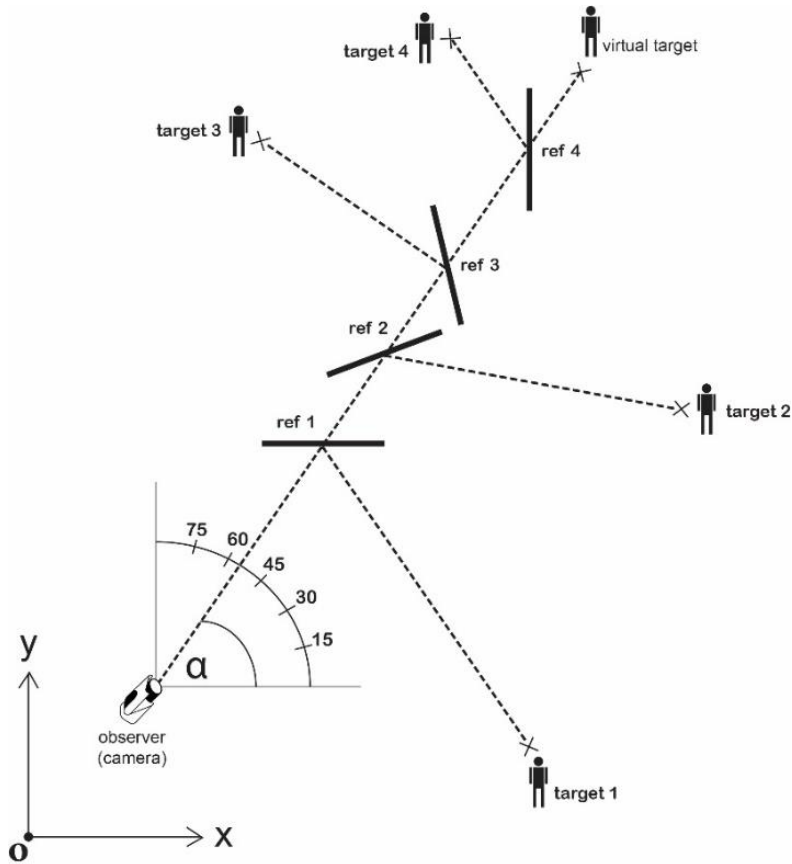

Figure 1. Localization of a person in NLOS assuming one-bounce reflection through a mirror using a camera posed at a known position in the world frame.

The key challenge is how to benefit from the limited information coming from beacons to achieve the localization. In this paper, we want to proof that with only one observer and a determined number of beacons, the target localization problem is able to solved, for an unknown map. Also a beacon placement analysis is done; the number of needed beacons and their locations are proposed. The problem is addressed for the limited size reflector.

## 2. Problem Formulation

Localizing a target after being reflected through a reflector depends on the reflector's pose and orientation, hence, the reflector should be localized. In a 2-D space, as shown in Fig. 2, in a world frame, a target's position, noted by $X_{tar} = (x_{tar}, y_{tar})^T$ is determined based on the observer and the reflector's poses, noted by $X_{obs} = (x_{obs}, y_{obs})^T$ and $X_{ref} = (x_{ref}, y_{ref})^T$ respectively, the reflector's orientation $(\theta_{ref})$ and the angle of the arrival of the target's observation $(\alpha)$, in which:

$$x_{tar} = x_{obs} + d_1 \cos(\alpha) + d_2 \cos(\alpha - 2\theta_{ref}) \tag{1a}$$

$$y_{tar} = y_{obs} + d_1 \sin(\alpha) + d_2 \sin(\alpha - 2\theta_{ref}) \tag{1b}$$

Where, $d_1$ is the distance between the observer and the reflector:

$$d_1 = \sqrt{(x_{obs} - x_{ref})^2 + (y_{obs} - y_{ref})^2} \tag{2}$$

And $d_2$ is the distance between the reflector and the target.

$$d_2 = \sqrt{(x_{tar} - x_{ref})^2 + (y_{tar} - y_{ref})^2} \tag{3}$$

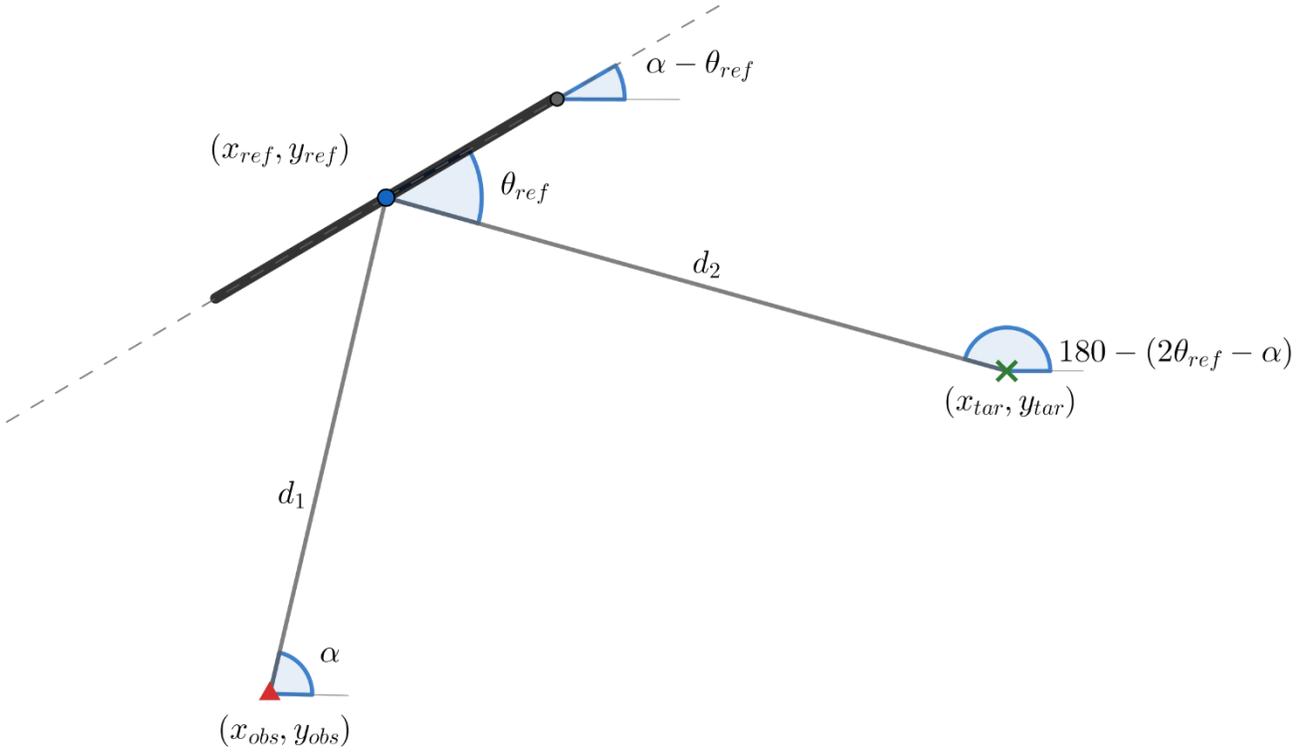

Figure 2. For NLOS environment, the position of the target depends on the reflector's pose and orientation, here, the notations of all components are shown in details.

With a total estimated distance from the observer to the target noted by $d_{total}$:

$$d_{total} = d_1 + d_2 \tag{4}$$

A relation between the object's size and the distance from the observer is introduced in [26, 27], in which the product of the size and the distance for an object is constant, it is the size constancy concept. It seems to be the same as the geometric projective of a picture on the image plane and from its size (the number of the occupied pixels), the distance of the object can be calculated. In other words, the distance of the object from the observer increases and decreases based on its size, when the product is calculated based on a reference size $(s_{rfr})$ at a reference distance $(d_{rfr})$.

$$s_{rfr}.d_{rfr} = s_{obj}.d_{total} = a \tag{5}$$

Where, $(s_{obj})$ is the size of the object seen by the observer.

The number of unknowns of this problem is five, in which two for the target's pose and three for the reflector. The target's pose depends on the reflector, for that, three independent variables must be determined, i.e. $(x_{ref}, y_{ref}, \theta_{ref})$.

Three equations can solve the problem; they can be extracted from a single beacon with known position:

$$x_{bea} = x_{obs} + d_1 \cos(\alpha_{bea}) + d_2 \cos(\alpha_{bea} - 2\theta_{ref}) \tag{6a}$$

$$y_{bea} = y_{obs} + d_1 \sin(\alpha_{bea}) + d_2 \sin(\alpha_{bea} - 2\theta_{ref}) \tag{6b}$$

$$d_{total}^{bea} = d_1^{bea} + d_2^{bea} \tag{6c}$$

But these equations are non-linear and must be linearized in order to localize the target.

## 3. Proposed Approach

Based on the results from [26], for a known position of the reflector, the reflector's orientation is determined using only 2 beacons, placed in the intersection area of all field of views of the observer for different reflector's orientation (see Fig. 3).

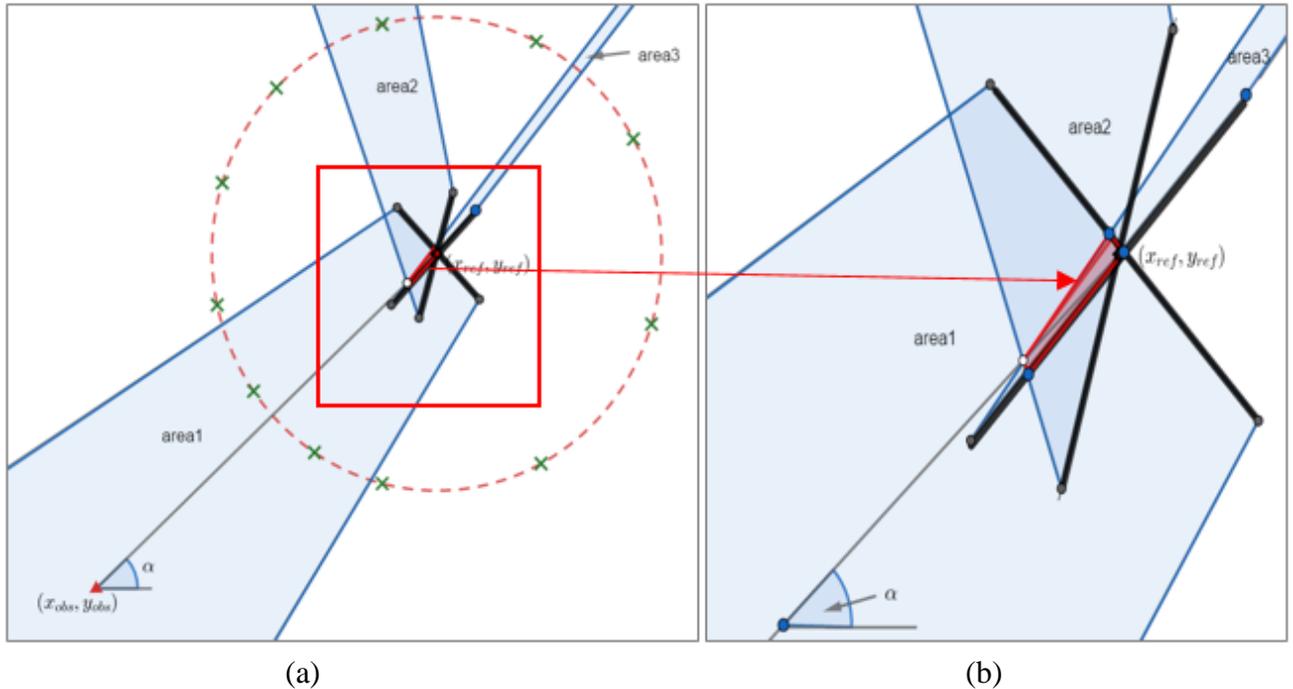

(a)  (b)

Figure 3. From [26], (a) an example of the intersection area of three field of views of the observer for a limited size reflector to put the beacon and ensure it to be seen by the observer. (b) Zoom In of the (a) to see clearly the desired area.

For an unknown reflector's position and orientation, the same work can be done; for the different reflector's positions along the line observer, two beacons must be placed in order to find the unknown reflector's orientation.

Here, the total estimated distance $d_{total}$ is divided into steps ($\Delta d$); for each step, the position of the reflector is known, and using the 2 beacons in the desired areas, its orientation is determined. The position of the reflector for each step is determined by:

$$d = d + i * \Delta d \tag{7}$$

Where $d : 0$ to $d_{total}$ and $\Delta d$ is the step size calculated in the later section. But for this method, the number of beacons grows; it depends on the total estimated distance, the size of the reflector and the threshold between the reflector and the line observation, i.e. $\varepsilon$ as shown in Fig. 4.

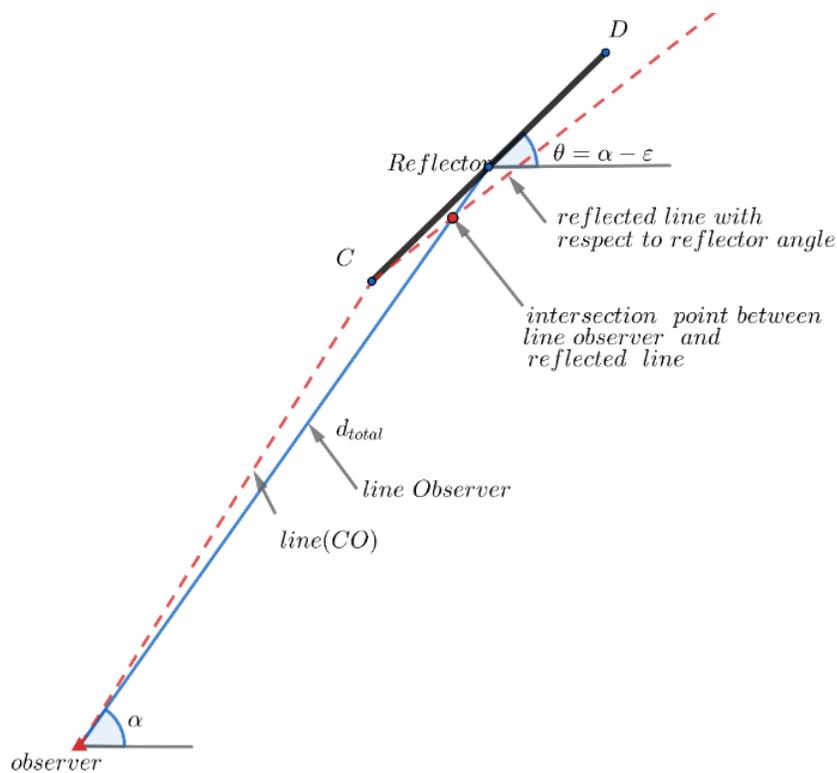

Figure 4. The red point is the intersection point between the reflected line passing through the edge of the reflector (point C) and the observation line.

The number of beacons needed to cover all these areas is:

$$N_{bea} = 2 * \frac{d_{total}}{step_{size}} \tag{8}$$

As shown in Fig. 4, for an angle of orientation of the reflector $\left(\theta_{ref}=\varepsilon\right)$ w.r.t the line observer, the line reflected from the observer line lying the observer with the edge of the reflector (point C) with respect to the reflector with angle $\left(\theta_{ref}=\alpha-\varepsilon\right)$ in the world frame, is determined by its angle $\left(\beta=\alpha-2\varepsilon\right)$ and the point C, calculated by:

$$x_c = d_{total}\cos(\alpha) - \frac{s}{2}\cos(\theta) \tag{9a}$$

$$y_c = d_{total}\sin(\alpha) - \frac{s}{2}\sin(\theta) \tag{9b}$$

The equation of this line should:

$$y = \tan(\alpha-2\varepsilon).x + y_C - x_C.\tan(\alpha-2\varepsilon) \tag{10}$$

The line observer is also determined by the observer point and one angle $(\alpha)$:

$$y = \tan(\alpha).x + y_R - x_R.\tan(\alpha) \tag{11}$$

With $x_R = d_{total}\cos(\alpha)$ and $y_R = d_{total}\sin(\alpha)$.

The intersection point between these 2 lines is determined by:

$y = y \Rightarrow$

$$\tan(\alpha-2\varepsilon).x + y_C - x_C.\tan(\alpha-2\varepsilon) = \tan(\alpha).x + y_R - x_R.\tan(\alpha) \tag{12}$$

Where

$$x_{inter1} = \frac{x_C.\tan(\alpha-2\varepsilon) - x_R.\tan(\alpha) + y_R - y_C}{\tan(\alpha-2\varepsilon) - \tan(\alpha)} \tag{13a}$$

$$y_{inter1} = \tan(\alpha).x_{inter} + y_R - x_R.\tan(\alpha) \tag{13b}$$

The step size is calculated by:

$$b = \Delta d = \sqrt{\left(x_{inter1} - x_R\right)^2 + \left(y_{inter1} - y_R\right)^2} \tag{14}$$

Now, the step size $(\Delta d)$ is determined, the position of the 2 beacons also must be determined. For that, as shown in Fig. 5, the beacons must be placed within the line of length b from the reflection point. It is the intersection between the reflector at angle 90 w.r.t to line observer, and the reflected line passing through the point C calculated using (10). The reflector has an angle of $90-\alpha$ w.r.t the global frame, and the point R. so its equation is:

$$y = \tan(90-\alpha).x + y_R - x_R.\tan(90-\alpha) \tag{15}$$

For that,

$y = y \Rightarrow$

$$\tan(\alpha-2\varepsilon).x + y_C - x_C.\tan(\alpha-2\varepsilon) = \tan(90-\alpha).x + y_R - x_R.\tan(90-\alpha) \tag{16}$$

It results in:

$$x_{inter2} = \frac{x_C.\tan(\alpha-2\varepsilon) - x_R.\tan(90-\alpha) + y_R - y_C}{\tan(\alpha-2\varepsilon) - \tan(90-\alpha)} \tag{17a}$$

$$y_{inter2} = \tan(90-\alpha).x_{inter2} + y_R - x_R.\tan(90-\alpha) \tag{17b}$$

Finally, the place of beacons is determined:

$$a/2 = \sqrt{(x_{inter2} - x_R)^2 + (y_{inter2} - y_R)^2} \tag{18}$$

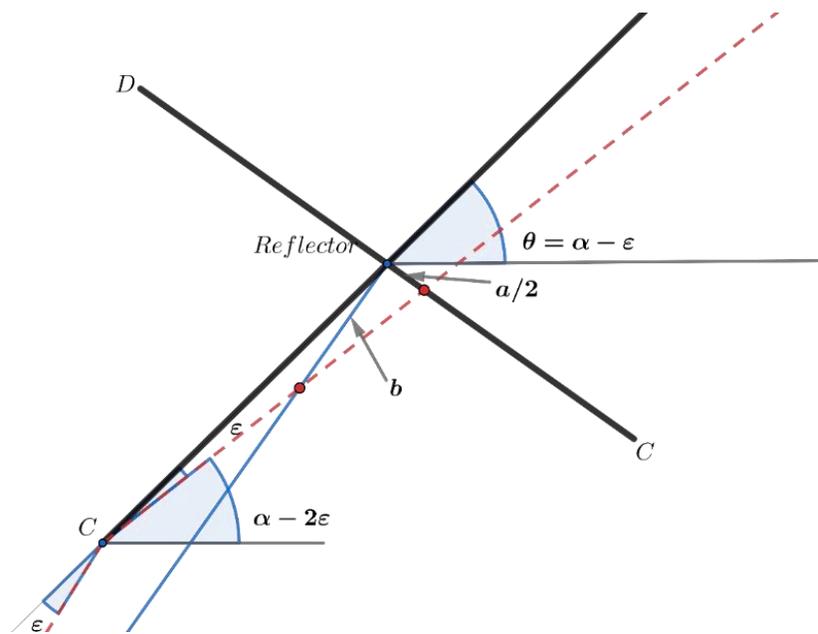

Figure 5. the non-moving beacon must be placed within the triangle rectangle area circumvented by the a/2 side and the b side.

An algorithm can be proposed for this method:

*Algorithm.1: SLTR*

01 **Input** $X_{obs}, S_{ref}, \varepsilon$
02 **For** $\alpha_{obs} = 0$ to $2\pi$ **do**
03      $\alpha_{tar} \leftarrow \alpha_{obs}$
04      $d_{total}(tar) \leftarrow d_{total}$
05      calculate $\Delta d$ and a
06 **EndFor**
07 **For** $d = 0: \Delta d : d_{total}$, $d_{new} = d_{old} + \Delta d$ **do**
08      place 2 beacons below the reflector, at the 2 sides of the observation's line
09      $\alpha_{obs} = \alpha_1$ to $\alpha_2$
10      $\alpha_{bea} \leftarrow \alpha_{obs}$ get the angle of the reflected beacon's signal
11      $d_{total}(bea) \leftarrow d_{total}$ get the distance from the beacon after reflection
12 Calculate $(D_{ref})$ using Eqs (),() and (), get the reflector's position and orientation
13 **EndFor**
14 **Output** $X_{tar}, (D_{ref})$

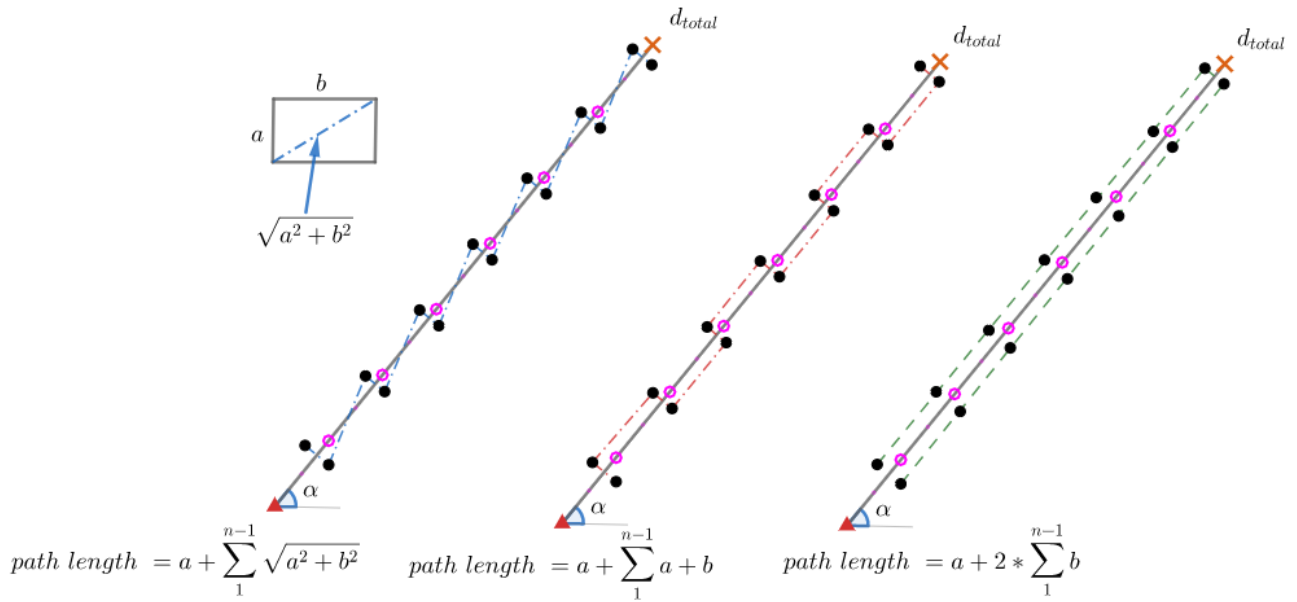

Figure 5. For a moving beacon, three paths are suggested here, from left to right: the saw-tooth path, the rectangular path and the linear path.

Due to the big number of needed beacons, this method cannot be applicable. Furthermore, it is imperative to reach the virtual target's position in order to place the beacons, i.e. to reach the last point of the possible position of the reflector along the observation line. So a moving beacon is proposed, in which it moves in a determined path to ensure that it is reflected by the reflector and seen by the observer, so the reflector (orientation and position) is determined first and the target is localized.

Many paths can be found, in Fig. 5, we suggest three paths, where one of them is the shortest path (based on the length):

Right) Path1-linear path: it is shown in green; the beacon moves from the first beacon's position, parallel to the line observer, and return from the other side. The total length of the path is:

$$P_{L1} = a + 2 * \sum_{1}^{n-1} b \tag{19}$$

Middle) Path2-rectangular path: it is shown in orange; the beacon moves from the first beacon's position in a rectangular form. The total length of the path is:

$$P_{L2} = a + \sum_{1}^{n-1} (a+b) \tag{20}$$

Left) Path3-saw-tooth path: it is shown in blue; the beacon moves from the first beacon's position, in saw-tooth form. The total length of the path is:

$$P_{L3} = a + \sum_{1}^{n-1} \left( a + \sqrt{a^2 + b^2} \right) \tag{21}$$

Comparing $P_{L1}, P_{L2}$ and $P_{L3}$, we see that for $b > a$ (I mean if the step size is bigger than the distance between the 2 beacons), then $a + b < \sqrt{a^2 + b^2} < 2*b \Rightarrow P_{L2} < P_{L3} < P_{L1}$. In conclusion, $P_{L2}$ is the shortest one between these three paths.

## 4. Beacon Motion Model and Particle Filter

In practice, two motion models are existing, the odometry-based and the velocity-based. Due to the different beacon's types (UGV or UAV), a velocity-based model is suggested in which the pose of the beacon can be calculated based on the velocity and the time elapsed. Here, we have to model the translation between 2 consecutive poses, without caring the rotation. The beacon moves from $X^{(t)} = \left( x_{bea}^{(t)}, y_{bea}^{(t)} \right)^T$ to $X^{(t+1)} = \left( x_{bea}^{(t+1)}, y_{bea}^{(t+1)} \right)^T$, so the translation is calculated by $t = d\left( X_{bea}^{(t)}, X_{bea}^{(t+1)} \right)$. The measured motion is given by the true motion corrupted with noise, for that, we have $trs_{true} = trs_{mea} + n(\sigma_{trs})$. The translational velocity at time $t$ is noted by $v_t$ and the goal is to compute the probability $p(X^{(t+1)} | X^{(t)}, v_t)$. Here, a particle filter can be used; random samples are generated from $p(X^{(t+1)} | X^{(t)}, v_t)$ for a known pose $X^{(t)}$ and fix velocity $v_t$. And a set of random pose $X^{(t+1)}$ are generated according to the distribution $p(X^{(t+1)} | X^{(t)}, v_t)$. Due to the error in the kinematic motion model, noises are added and they are used to generate the new samples of the beacon's pose $X^{(t+1)}$.

## 5. Simulation Results

Our measurements are the angle of arrival of the target's signal and the distance estimated from the target to the observer. for that, the errors can be modeled as Gaussian noises zero means with $v(\alpha)$ variance and $v(d_{total})$ variance for both the angle of arrival and the estimated distance respectively. It is important to say that for a perfect camera, the distance can be estimated without error, i.e. the size is calculated precisely. Here, the noise in the estimation of the distance is due to the concavity or convexity of the mirror, it can be not ideal in which the noise occurs.

The intersection area of the field of views of the observer for different reflector's orientations is determined. It is a triangle rectangle in R (the intersection point between the observation line and the reflector) with 2 sides, a/2 and b as shown in the Fig. 6. Because of noisy measurements, this triangle is shifted and the area is varied due to the SD values.

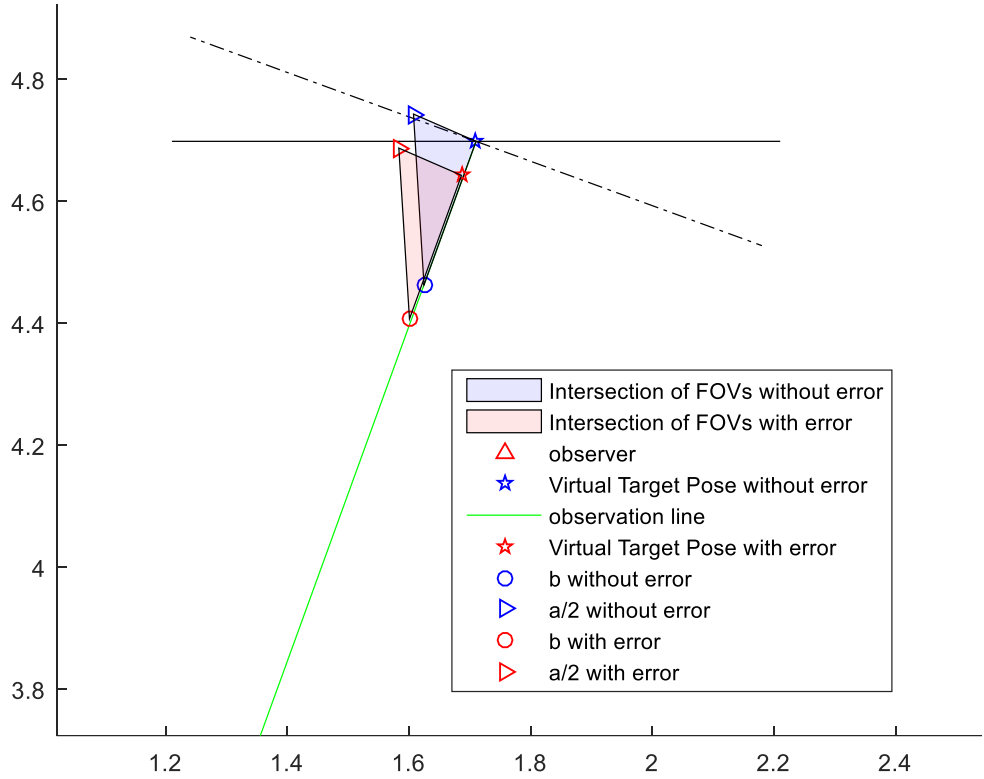

Figure 6. From simulation, the lossless intersection area is shown in blue, and after adding noises to the measurements, the noisy area should in red, it is shifted from the ideal case.

Table 1
The RMSEs of both reflector and target after adding the standard deviations for the measurements

| $v(\alpha)$ | $v(d_{total})$ | RMSE(ref) | RMSE(tar) |
|---|---|---|---|
| 1.73 | 0.16 | 0.036 | 0.10 |
| 2.60 | 0.29 | 0.09 | 0.12 |
| 1.46 | 0.41 | 0.23 | 0.15 |

To show the validity of our approach, the RMSEs of the target and reflector are calculated based on random standard deviations up to 10% of the real value of the measurements (the total estimated distance is 5 and the angle of arrival is 70), for 200 iterations each run. The results are shown in Table 1.

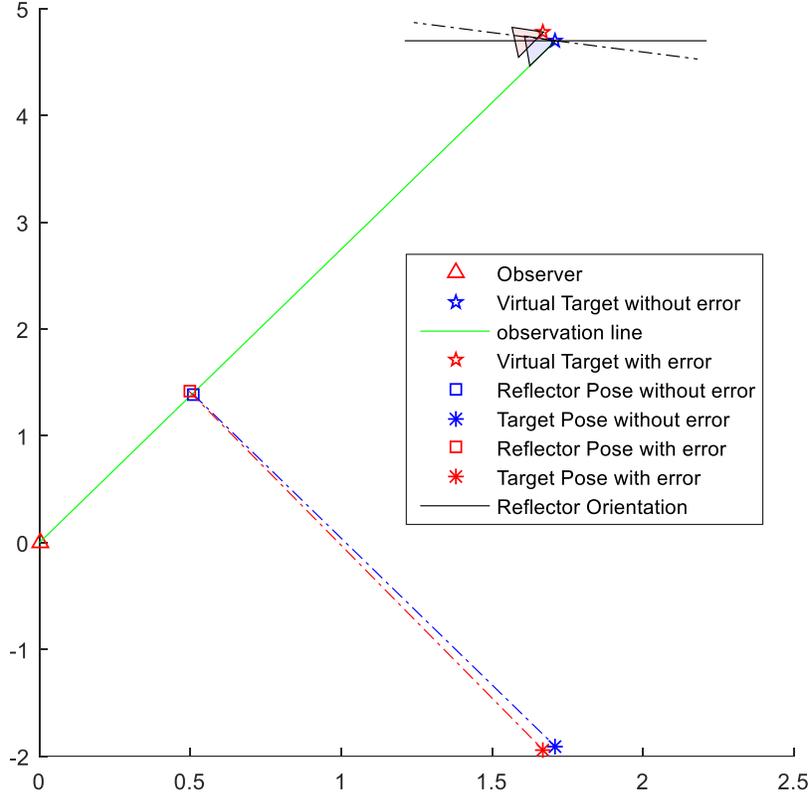

Figure 7. Based on the beacon placed within the desired area, the target and reflector are localized. The true positions of them are shown in blue (square for the reflector and star for the target), and the noisy positions are shown in red.

## 6. Experimental Results

In order to test the proposed approach, a mirror space is considered (See Fig. 8) and a set of experiments are performed. The target was tested with different positions of the target and different mirror's orientations and positions. The noises of the measurements (the angle of coming signal and the estimated distance between the target and the camera) are modeled based on the reflector's specs, the camera's resolution and the exact positions of the components into the searching area.

Setup of the experiments:

- An 80x80 $cm^2$ plane is formed with uniform holes.
- The experiments are done using only a single receiver as an observer, it is a camera with 2 MP resolution (1088*1920).

- The camera has a fixed position, but free direction, i.e. it can scan in all directions.
- The mirror is chosen to be a perfect reflector with a smooth surface as much as possible.
- The target is an object of known size constancy value, i.e. it has a known size at a reference distance.

Before starting:

- The camera must be calibrated.
- The reflector, the camera and the object must be checked to be 90° with the plane to prevent the skewed object's image.

Steps of the work:

- A target is placed at a known position, facing the mirror, to be seen by the camera directed at the object's center, and the length of the object is determined by the pixels covering its image in the picture.
- After getting the estimated distance and the angle of arrival of the observation, a beacon with known position and size, is placed at the required area; the image of the beacon is seen by the camera but in NLOS, with different target's angle of arrival.
- The true beacon's position is known, and its image is calculated, so the orientation and the position of the reflector is determined using (1a), (1b) and (4).
- Knowing the reflector's position and orientation, the target is localized using Eqs (6a), (6b) and (6c).
- This scenario is repeated for n times, in order to model the error from real measurements.

Table 2
The necessary components used in the experimental work and their sizes.

|  | Size (cm$^2$) | Type | Size constancy value |
|---|---|---|---|
| Reflector | 18.2*13.25 | Planar mirror | --- |
| Target | 15.25*9.9 | Rectangular shape | 15.25 |
| Searching area | 80*80 | Square plane | --- |
| Beacon | 4.2*2.9 | Rectangular shape | 4.2 |

It is important to say that the size constancy value of the objects in table.2 is calculated based on 1 m distance for both the target and the beacon, and it can be transformed to the pixel world, and each size is calculated based on the occupied pixels on the image plane. For example, for the target with size 15.25 cm at 1 m, its size constancy is 15.25, but assuming that we work in the world image, its size constancy value must be 229 (it occupies 229 pixels in length at 1 m distance). The same work is done for the beacon.

Finally, the goal is to localize the target, after determining the orientation and position of the reflector using beacons. The total noises are deduced and modeled.

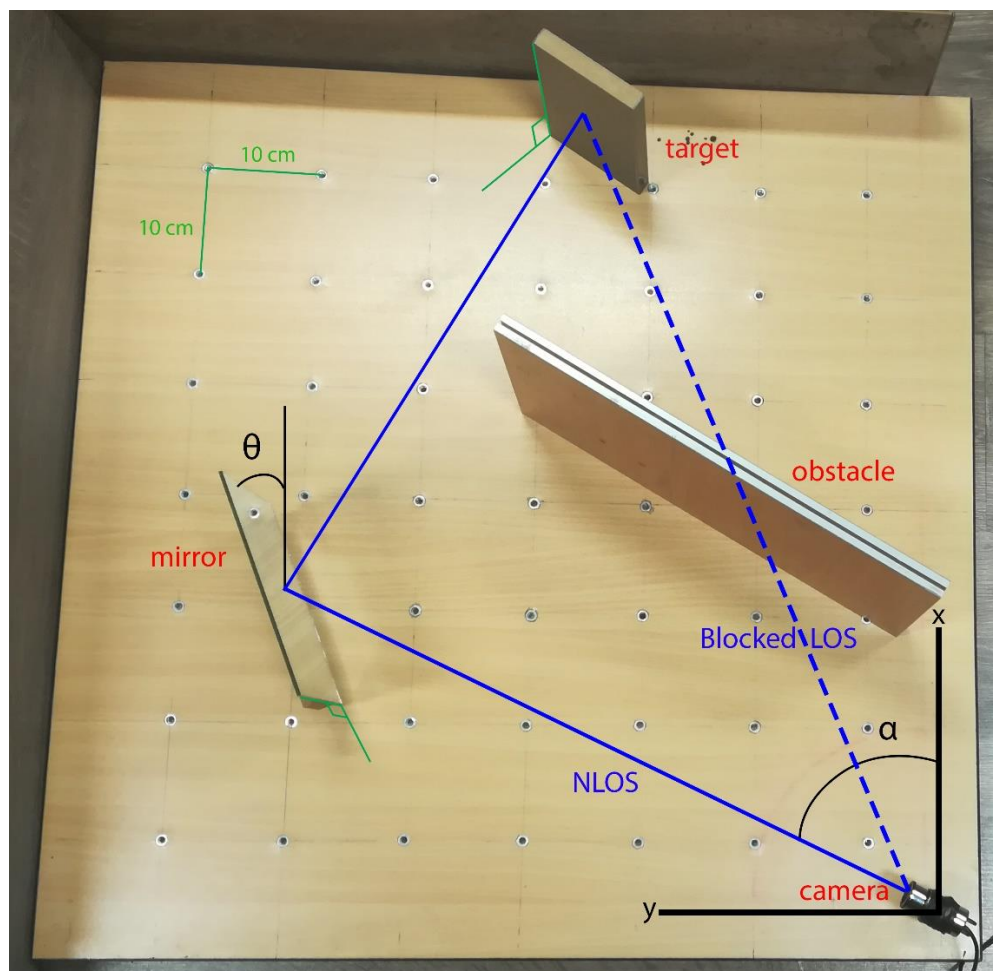

Figure 8. The real work is implemented in the mirror space, the target (grey rectangle) is seen indirectly by a camera fixed at the corner of the plane through the mirror. From the orientation of the camera and the size of the object projected at the image frame, the angle of arrival of the observation and the estimated distance from target to the camera are determined respectively.

Table 3
The RMSE of the target localization and the measured data vs the real data for different experiments (all the positions and distances are in cm, and the orientations are in degrees).

| One bounce reflection | | EXP1 | EXP2 | EXP3 |
|---|---|---|---|---|
| Real data | $X_{ref}$ | (27.5,65) | (15.5,34) | (16,34.5) |
| | $\theta_{ref}$ | 38 | 0 | 38.3 |
| | $X_{tar}$ | (55,70) | (23.5,16.4) | (65.5,42.7) |
| | $d_{total}(tar)$ | 98.59 | 56.7 | 79.68 |
| | $\alpha(tar)$ | 65 | 63 | 62.3 |
| | $X_{bea}$ | (32, 62.5) | (18.2,0) | (17,33.5) |
| | $\alpha(bea)$ | 75 | 68.5 | 68.2 |
| | $d_{total}(bea)$ | 75.7 | 67.64 | 40.8 |
| Measured data | $X_{ref}$ | (29.3,62.5) | (15,34.8) | (14.6,36.2) |
| | $\theta_{ref}$ | 35.42 | 1.7 | 41.04 |
| | $X_{tar}$ | (52,65) | (23.1,17.08) | (60.5,48,6) |
| RMSE(tar) | | 5.83 | 3.79 | 7.7337 |

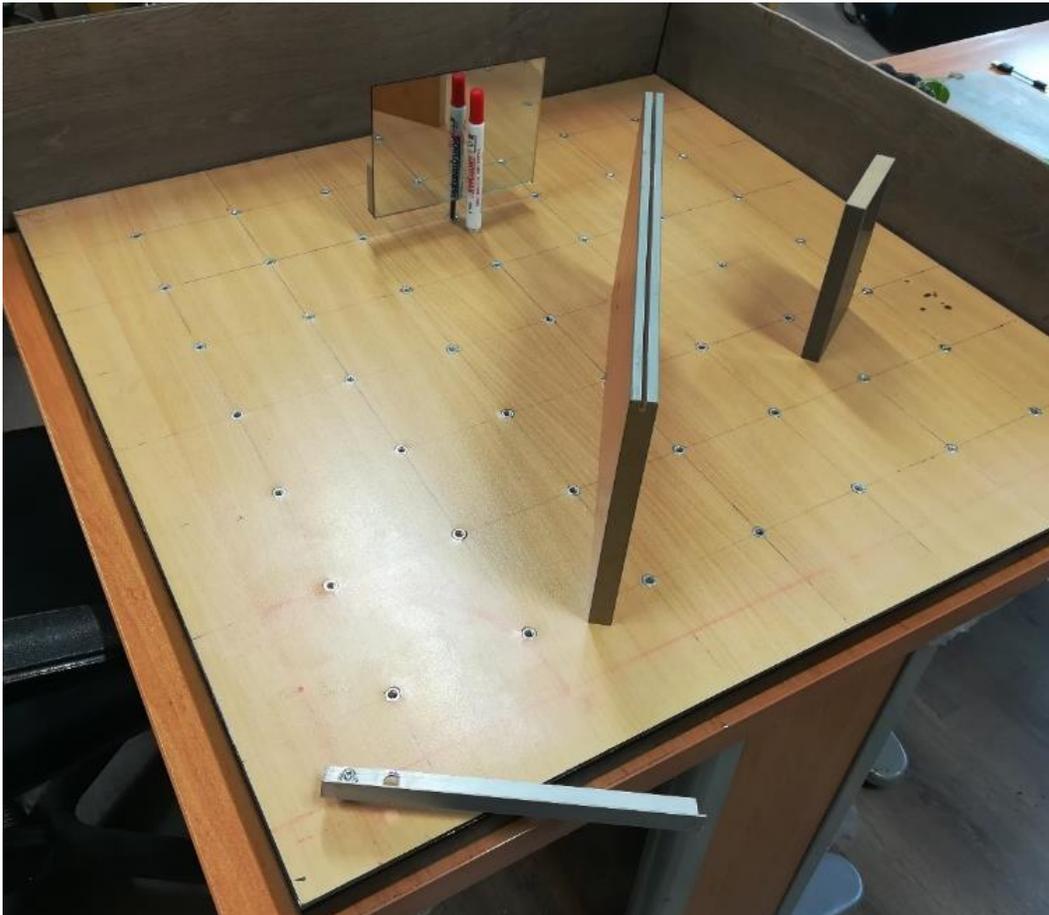

Figure 10. the reflector's position and orientation (i.e. the mirror) is estimated using a single beacon (the pencil) placed at the desired area.

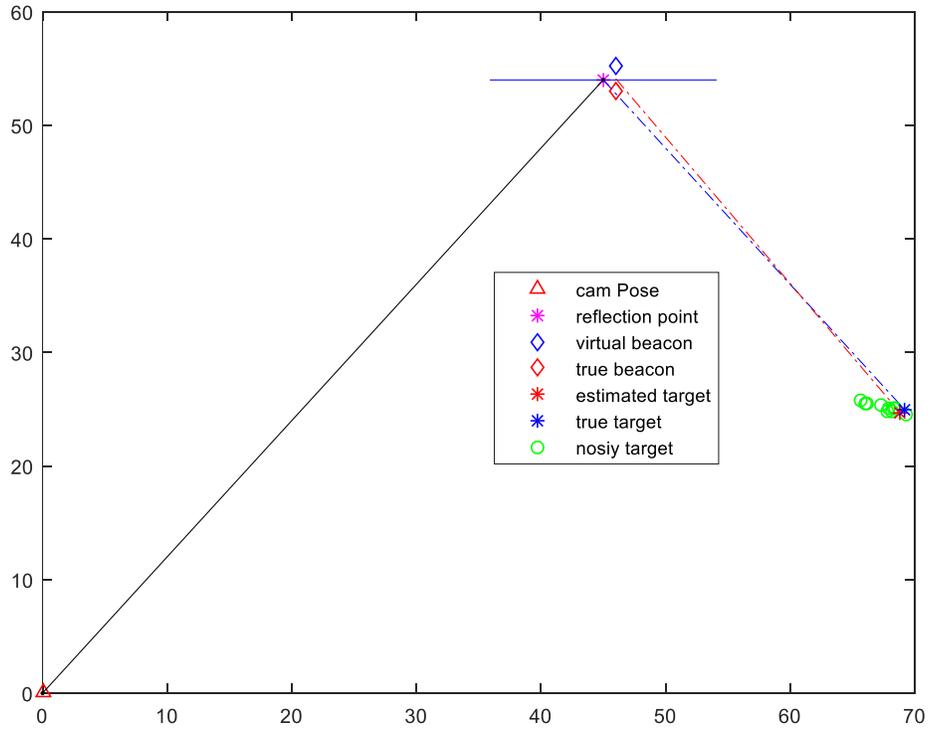

Figure 11. the true and estimated position of the target is determined based on the beacon's real and virtual position.

## 7. Discussion

Based on the experimental and simulation results, the proposed approach is applicable for the mirror space and the other mentioned spaces (RF and sound and…), but here, the conditions must be changed. For example, in our work, the mirror was $1/5^{th}$ the total distance, but usually, in RF world, the total distance is widely bigger than the reflector's size (let's say s/d=1/1000); if this approach is valid for this ratio? The intersection area founded for placing a beacon is the triangle rectangle on the reflector pose at $(d = d_{total})$ . And their sides are described in (14) and (18). By expanding these equations, the total distance $(d_{total})$ can be eliminated, and the sides of the intersection areas are dependent on the size of the reflector, the angle of the coming observation $(\alpha)$ and the maximum reached reflector's orientation $(\theta_{ref} = \varepsilon)$. So the intersection area is independent of the total distance.

## 8. Conclusion

In this paper we addressed the target localization issue, after reflection from a reflector, in NLOS condition and solved it using beacons. Real experiments are done in a mirror space, in which the observer is a camera and the reflector is a planar mirror. Hence, the target is an object snapped by a camera and from its size that is projected at the image plane of the camera, the distance from target to the camera is estimated using size constancy concept. This is a typical application of patrolling areas and localize persons or objects in the desired areas. Although the traditional localization methods can solve most of the cases, especially when they are mixed or combined, but our results show that using only one observer and a determined number of beacons, the impossible target localization cases are able to be solved. Furthermore, we have proposed an algorithm to determine the best positions for the beacons, to minimize the localization error. The approach has been formulated and tested in simulated environment; in conclusion, the experimental results match the theoretical results.

In the future, this work can be extended to other spaces such as the sound source localization and RFsource localization. Also, the same problem can be addressed for 3-D space.

**References**

[1] M. Alain and B. H. Irene, "Indoor positioning through fingerprinting technics: How many beacons should be deployed and where?," in *Wireless Personal Multimedia Communications (WPMC), 2017 20th International Symposium on*, 2017, pp. 522-528: IEEE.
[2] P. Kriz, F. Maly, and T. Kozel, "Improving indoor localization using bluetooth low energy beacons," *Mobile Information Systems,* vol. 2016, 2016.
[3] Z. Ma and K. Ho, "TOA localization in the presence of random sensor position errors," in *Acoustics, Speech and Signal Processing (ICASSP), 2011 IEEE International Conference on*, 2011, pp. 2468-2471: IEEE.
[4] S. Lee, W. Lee, and K. You, "TDOA based UAV localization using dual-EKF algorithm," in *Control and Automation*: Springer, 2009, pp. 47-54.
[5] X. Song, P. Willett, and S. Zhou, "Target localization with NLOS circularly reflected AoAs," in *2011 IEEE International Conference on Acoustics, Speech and Signal Processing (ICASSP)*, 2011, pp. 2472-2475: IEEE.

[6] S. C. Ergen, H. S. Tetikol, M. Kontik, R. Sevlian, R. Rajagopal, and P. Varaiya, "RSSI-fingerprinting-based mobile phone localization with route constraints," *IEEE Transactions on Vehicular Technology,* vol. 63, no. 1, pp. 423-428, 2014.

[7] G. Ding, Z. Tan, L. Zhang, Z. Zhang, and J. Zhang, "Hybrid TOA/AOA cooperative localization in non-line-of-sight environments," in *Vehicular Technology Conference (VTC Spring), 2012 IEEE 75th*, 2012, pp. 1-5: IEEE.

[8] F. Dai, Y. Liu, and L. Chen, "A Hybrid Localization Algorithm for Improving Accuracy Based on RSSI/AOA in Wireless Network," in *Computer Science & Service System (CSSS), 2012 International Conference on*, 2012, pp. 631-634: IEEE.

[9] Z. Wang, *Multi-node TOA-DOA cooperative LOS-NLOS localization: enabling high accuracy and reliability*. Michigan Technological University, 2010.

[10] L. Ren *et al.*, "De2: Localization with the rotating rss using a single beacon," in *Ubiquitous Intelligence and Computing, 2013 IEEE 10th International Conference on and 10th International Conference on Autonomic and Trusted Computing (UIC/ATC)*, 2013, pp. 9-16: IEEE.

[11] J.-R. Jiang, C.-M. Lin, F.-Y. Lin, and S.-T. Huang, "ALRD: AoA localization with RSSI differences of directional antennas for wireless sensor networks," *International Journal of Distributed Sensor Networks,* vol. 2013, 2013.

[12] R. Allen, N. MacMillan, D. Marinakis, R. I. Nishat, R. Rahman, and S. Whitesides, "The range beacon placement problem for robot navigation," in *Computer and Robot Vision (CRV), 2014 Canadian Conference on*, 2014, pp. 151-158: IEEE.

[13] J. Yick, A. Bharathidasan, G. Pasternack, B. Mukherjee, and D. Ghosal, "Optimizing placement of beacons and data loggers in a sensor network-a case study," in *Wireless Communications and Networking Conference, 2004. WCNC. 2004 IEEE*, 2004, vol. 4, pp. 2486-2491: IEEE.

[14] C. Wan, A. Mita, and S. Xue, "Non-line-of-sight beacon identification for sensor localization," *International Journal of Distributed Sensor Networks,* vol. 8, no. 8, p. 459590, 2012.

[15] J. J. Leonard and H. F. Durrant-Whyte, "Mobile robot localization by tracking geometric beacons," *IEEE Transactions on robotics and Automation,* vol. 7, no. 3, pp. 376-382, 1991.

[16] M. L. Sichitiu and V. Ramadurai, "Localization of wireless sensor networks with a mobile beacon," *MASS,* vol. 4, pp. 174-183, 2004.

[17] C. Schaff, D. Yunis, A. Chakrabarti, and M. R. Walter, "Jointly optimizing placement and inference for beacon-based localization," *arXiv preprint arXiv:1703.08612,* 2017.

[18] N. Rajagopal, S. Chayapathy, B. Sinopoli, and A. Rowe, "Beacon placement for range-based indoor localization," in *Indoor Positioning and Indoor Navigation (IPIN), 2016 International Conference on*, 2016, pp. 1-8: IEEE.

[19] C. Sha and R.-c. Wang, "A type of localization method using mobile beacons based on spiral-like moving path for wireless sensor networks," *International Journal of Distributed Sensor Networks,* vol. 9, no. 8, p. 404568, 2013.

[20] B. Xiao, H. Chen, and S. Zhou, "Distributed localization using a moving beacon in wireless sensor networks," *IEEE Transactions on Parallel and Distributed Systems,* vol. 19, no. 5, pp. 587-600, 2008.

[21] J.-M. Valin, F. Michaud, J. Rouat, and D. Létourneau, "Robust sound source localization using a microphone array on a mobile robot," in *Proceedings 2003 IEEE/RSJ International Conference on Intelligent Robots and Systems (IROS 2003)(Cat. No. 03CH37453)*, 2003, vol. 2, pp. 1228-1233: IEEE.

[22] I. An, M. Son, D. Manocha, and S.-e. Yoon, "Reflection-Aware Sound Source Localization," in *2018 IEEE International Conference on Robotics and Automation (ICRA)*, 2018, pp. 66-73: IEEE.

[23] S. Zorn, R. Rose, A. Goetz, and R. Weigel, "A novel technique for mobile phone localization for search and rescue applications," in *Indoor Positioning and Indoor Navigation (IPIN), 2010 International Conference on*, 2010, pp. 1-4: IEEE.

[24] R. Severino and M. Alves, "Engineering a search and rescue application with a wireless sensor network-based localization mechanism," in *World of Wireless, Mobile and Multimedia Networks, 2007. WoWMoM 2007. IEEE International Symposium on a*, 2007, pp. 1-4: IEEE.
[25] S. Savarese, M. Chen, and P. Perona, "Local shape from mirror reflections," *International Journal of Computer Vision,* vol. 64, no. 1, pp. 31-67, 2005.
[26] M. Fares, H. Moradi, M. Shahabadi, and S. Nasiri, "Target Localization in NLOS Condition Using Beacons to Determine Reflector Orientation," in *2018 6th RSI International Conference on Robotics and Mechatronics (IcRoM)*, 2018, pp. 178-182: IEEE.
[27] A. Higashiyama and K. Shimono, "Mirror vision: Perceived size and perceived distance of virtual images," *Attention, Perception, & Psychophysics,* vol. 66, no. 4, pp. 679-691, 2004.